\documentclass[useAMS,usenatbib]{mn2e} 

\usepackage{graphicx}      
\usepackage{txfonts} 
\usepackage[authoryear]{natbib}
\usepackage{subfigure}

\newcommand{\teff}{\ifmmode T_{\rm eff} \else T$_{\mathrm{eff}}$\fi}
\newcommand{\lL}{\ifmmode \log \frac{L}{L_{\odot}} \else $\log \frac{L}{L_{\odot}}$\fi}

\newcommand{\myr}{M$_{\odot}$ yr$^{-1}$}
\newcommand{\vsini}{$v {\mathrm sin}i$}

\newcommand{\kms}{km~s$^{-1}$}
\newcommand{\msun}{\ifmmode M_{\odot} \else M$_{\odot}$\fi}
\newcommand{\hei}{\hbox{He$\;${\sc i}}}
\newcommand{\heii}{\hbox{He$\;${\sc ii}}}
\newcommand{\niii}{\hbox{N$\;${\sc iii}}}
\newcommand{\oiii}{\hbox{O$\;${\sc iii}}}
\newcommand{\civ}{\hbox{C$\;${\sc iv}}}
\newcommand{\hal}{\hbox{H${\alpha}$}}

\def\sn{\hbox{S/N}}
\def\ptt{\hbox{$10^{-4} I_{\rm c}$}}
\def\chisqr{\hbox{$\chi^2_{\rm r}$}}

\def\tori{\hbox{$\theta^1$~Ori~C}}

\title[Detection of a magnetic field on HD~108]
{Detection of a magnetic field on HD~108: clues to extreme magnetic
  braking and the Of?p phenomenon\thanks{Based on observations
    obtained at the T\'elescope Bernard Lyot (TBL) and at the
    Canada-France-Hawaii Telescope (CFHT).  CFHT is operated by the
    National Research Council of Canada, the Institut National des
    Sciences de l'Univers of the Centre National de la Recherche
    Scientifique of France (INSU/CNRS), and the University of Hawaii,
    while TBL is operated by CNRS/INSU.}}

\author[Martins et al.]
{\vspace{0.2cm}
F. Martins$^{1}$, J.-F. Donati$^{2}$, W.L.F. Marcolino$^{3,4}$, J.-C. Bouret$^{3,5}$, G.A. Wade$^{6}$ \\
\vspace{0.1cm}
{\LARGE\rm C. Escolano$^{3}$, I.D.~Howarth$^{7}$ and the MiMeS Collaboration } \\ 
$^{1}$GRAAL--UMR5024, CNRS \& Universit\'e Montpellier II, Place Eug\`ene Bataillon, F--34095, Montpellier Cedex 05, France \\
$^{2}$LATT--UMR 5572, CNRS \& Univ.\ de Toulouse, 14 Av.\ E.~Belin, F--31400 Toulouse, France\\
$^{3}$LAM--UMR 6110, CNRS \& Universit\'e de Provence, rue Fr\'ed\'eric Joliot-Curie, F-13388, Marseille Cedex 13, France \\ 
$^{4}$Observat\'orio Nacional-MCT, Rua Jos\'e Cristino, 77, CEP 20921-400,S\~ao Crist\'ov\~ao, Rio de Janeiro, Brasil \\
$^{5}$NASA/GSFC, Code 665, Greenbelt, MD 20771, USA \\
$^{6}$Department of Physics, Royal Military College of Canada, PO Box 17000, Station Forces, Kingston, ON K7K 7B4, Canada \\
$^{7}$ Dept.\ of Physics and Astronomy, University College London,  Gower Street, London WC1E6BT, UK\\
} 
\begin{document} 

\date{Accepted ... Received ... in original form ...} 
 
\pagerange{\pageref{firstpage}--\pageref{lastpage}} \pubyear{2010} 
 
\maketitle 
 
\label{firstpage} 

\begin{abstract} 

We report the detection of a magnetic field on the Of?p star
HD~108. Spectropolarimetric observations conducted in 2007, 2008 and
2009 respectively with NARVAL@TBL and ESPaDOnS@CFHT reveal a clear
Zeeman signature in the average Stokes V profile, stable on timescales
of days to months and slowly increasing in amplitude on timescales of
years.  We speculate that this timescale is the same as that on which
H$\alpha$ emission is varying and is equal to the rotation period of
the star. The corresponding longitudinal magnetic field, measured
during each of the three seasons, increases slowly from 100 to 150~G,
implying that the polar strength of the putatively-dipolar large-scale
magnetic field of HD~108 is at least 0.5~kG and most likely of the
order of 1--2 kG.

The stellar and wind properties are derived through a quantitative
spectroscopic analysis with the code CMFGEN. The effective temperature
is difficult to constrain because of the unusually strong He~{\sc i}~$\lambda$~
4471 and He~{\sc i}~$\lambda$~5876 lines. Values in the range
33000--37000 K are preferred. A mass loss rate of about $10^{-7}$
\myr\ (with a clumping factor f=0.01) and a wind terminal velocity of
2000 \kms\ are derived. The wind confinement parameter $\eta_{\star}$
is larger than 100, implying that the wind of HD~108 is magnetically
confined.

Stochastic short-term variability is observed in the wind-sensitive
lines but not in the photospheric lines, excluding the presence of
pulsations. Material infall in the confined wind is the most likely
origin for lines formed in the inner wind. Wind--clumping also
probably causes part of the H$\alpha$ variability. The projected
rotational velocity of HD~108 is lower than 50 \kms, consistent with
the spectroscopic and photometric variation timescales of a few
decades. Overall, HD~108 is very similar to the magnetic O star
HD~191612 except for an even slower rotation.

\end{abstract} 
 

\section{INTRODUCTION}  
\label{s_intro}     

Very little is known about magnetic fields of O stars and very few O
stars are yet known as magnetic; preliminary results
\citep[e.g.,][]{donati02,donati06,bouret08,petit08} suggest that most
magnetic O stars could be the high-mass equivalent of magnetic
chemically peculiar A and B stars (the so-called Ap and Bp stars) with
fields of primordial origin, i.e., fossil remnants from the formation
stage \citep{dl09}.  They are also suspected to be progenitors of
highly-magnetic neutron stars and magnetars
\citep[e.g.][]{petit08,Ferrario08}, although alternative explanations
exist \citep{davies09}.

Whenever known, rotation rates of magnetic O stars are significantly smaller 
than non-magnetic O stars of similar spectral types, further strengthening the 
analogy with the lower-mass Ap and Bp stars.  Evolutionary models of massive 
stars including magnetic fields also suggest that the internal rotation is 
strongly affected by the presence of magnetic fields, enforcing in particular 
mostly solid-body rotation profiles throughout the bulk of the outer radiative 
zone \citep{Maeder03,Maeder04}.  Last but not least, magnetic fields are 
potentially capable of affecting the powerful radiative winds of O stars 
\citep{bm97} and, in some extreme cases, of fully dissipating their angular 
momentum through wind magnetic braking \citep{ud09}.  
Lives of magnetic O stars are therefore expected to differ significantly from 
those of their non-magnetic counterparts.  

In this respect, the prototypical cases of \tori\ \citep{donati02} and HD~191612 
\citep{donati06} are very interesting.  
The magnetic field of \tori\ (about 1.1~kG tilted at about 45\degr\ to the rotation 
axis) is strong enough to deflect the wind flows from both magnetic hemispheres 
towards the magnetic equator where they collide, generating a strong shock, a very 
hot (coronal-like) post-shock region and a denser cooling disc (more or less confined 
to the magnetic equatorial plane).  
Emission in Balmer lines exhibits very strong rotational modulation, the amount of 
emission being directly related to the geometry of the cooling disc as seen from the 
Earth: strongest when the disc is seen pole-on and weakest when the disc is seen 
edge-on \citep[e.g.,][]{donati02}.  
The presence of magnetic fields and the very similar \hal\ modulation reported 
for HD~191612 suggests that this star is analogous to \tori\ \citep{donati06}, but 
more evolved and much more slowly rotating (538~d vs.\ 15.4~d for \tori) as a 
likely result of wind magnetic braking.  Recent (yet unpublished) data 
(Wade et al, in preparation) fully confirms the initial conclusions and the 
remarkable similarity between HD~191612 and \tori.  

Exploring further the potential impact of magnetic fields on the life
of massive stars requires new magnetic O stars to be identified. From
the example of \tori\ and HD~191612, those exhibiting periodically
varying \hal\ emission are likely to be very good candidates. In
particular, HD~108 is interesting. It is a member of the Of?p class
composed of the Galactic stars HD~191612, HD148937, NGC~1624--2 and CPD -28$\deg$ 2561, the latter two recently discovered by \citet{walborn10}. This class is principally defined by emission in the C {\sc iii} 4647--4650 lines at least as strong as in the neighbouring N {\sc iii} 4630--4640 lines. Secondary classification criteria include sharp emission or P-Cygni profiles in He {\sc i} and Balmer lines.  As \tori\ and HD~191612, HD~108
shows spectral type variations (from 04 to 08) due to changes in the
strength of its He~{\sc i} lines. They (and the Balmer lines too)
fluctuate between pure absorption and P--Cygni profiles. These line
morphologies define ``low'' and ``high'' states respectively. The
changes are correlated to photometric variations: HD~108 is brighter
when the emission components are stronger, again similarly to
HD~191612. They seem to occur with a period of a few decades
\citep{naze01,barannikov07} compared to 538 days for HD~191612
\citep{walborn04,howarth07}. If the analogy with \tori\ and HD~191612
holds, HD~108 could thus potentially show up as a rather extreme case
of wind magnetic braking. Note that a marginal detection based on a single exposure was reported by \citet{hubrig08} for the third member of the Galactic Of?p class, HD148937 \citep[but see][for a discussion of the significance of the detection]{silv09}. Further observations are required to confirm this detection.

To investigate this issue, we initiated a spectropolarimetric
monitoring program of HD~108 using NARVAL at the 2~m T\'elescope
Bernard Lyot (TBL) in the French Pyrenees and later on ESPaDOnS at the
3.6~m Canada-France-Hawaii Telescope (CFHT) atop Mauna Kea,
Hawaii\footnote{This effort is part of the MiMeS (Magnetism of Massive
  Stars) international research program aimed at characterizing
  magnetic fields in upper main sequence stars.}.  Reports indicate
that HD~108 has just passed the epoch of minimum \hal\ emission
(Naz\'e et al, in preparation), suggesting that longitudinal magnetic
fields are also close to their minimum value and slowly rising again
(provided the analogy with \tori\ and HD~191612 holds).

We report here the first results after 3 observing seasons (2007 to 2009).  
In particular, we report the detection of a longitudinal magnetic field 
of about 100~G at the surface of HD~108, suggesting that the analogy with 
\tori\ and HD~191612 is likely.  We also carry out a detailed spectral 
analysis to obtain updated constraints on the parameters of HD~108.  
We finally discuss briefly the implications of our results for 
understanding of how magnetic fields affect the lives of massive stars.

\section{OBSERVATIONS}
\label{obs}

Spectropolarimetric observations of HD~108 were collected with NARVAL@TBL and ESPaDOnS@CFHT 
in 2007 October, 2008 October and 2009 July to October.  Altogether, 110 sequences (24 in 2007, 
27 in 2008 and 59 in 2009) were obtained in multiple runs, each sequence consisting of four 
individual subexposures taken in different polarimeter configurations.  

From each set of four subexposures we derive a mean Stokes $V$ spectrum
following the procedure of \citet{Donati97}, ensuring in particular that all
spurious signatures are removed at first order.
Null polarization spectra (labeled $N$) are calculated by combining the four 
subexposures in such a way that polarization cancels out, allowing us 
to check that no spurious signals are present in the data
\citep[see][for more details on how $N$ is defined]{Donati97}.
All frames were
processed using Libre~ESpRIT \citep{Donati97}, a fully automatic reduction package 
installed at TBL and CFHT for optimal extraction of NARVAL and ESPaDOnS spectra.  
The peak signal-to-noise ratios per 2.6~\kms\ velocity bin range from 140 to 1100, 
depending on the instrument, on the exposure time and on weather conditions 
(see Table~\ref{tab:log} for a complete log).

\begin{table*}
\caption[]{Journal of observations.  Columns 1--6 list the
date, the instrument used, the range of heliocentric Julian dates, the range of UT times,
the number of sequences and the exposure time per individual sequence
subexposure, and the range of peak signal to noise ratio
(per 2.6~\kms\ velocity bin), for each night of observation.
Column 7 lists the rms noise level (relative to
the unpolarized continuum level and per 7.2~\kms\ velocity bin) in
the circular polarization profile produced by Least-Squares
Deconvolution once averaged over the whole night. Column 8 gives the longitudinal field strength (and the corresponding 1$\sigma$ error bar) determined from the night averaged polarization spectrum. } 
\begin{tabular}{cccccccc}
\hline
Date & Instrument & HJD          & UT      & $t_{\rm exp}$ & \sn\  & $\sigma_{\rm LSD}$ & Bz\\
     &            & (2,454,000+) & (h:m:s) &   (s)         &       &   (\ptt)         & (G) \\
\hline
2007 Oct 15    & NARVAL   & 389.38926--389.47736 & 21:14:22--23:21:14 & $3\times4\times900$ & 500--590 & 2.2 & -94.2$\pm$54.3\\ 
2007 Oct 16/17 & NARVAL   & 390.48564--390.57375 & 23:33:12--01:40:05 & $3\times4\times900$ & 350--550 & 2.6 & -51.3$\pm$63.8\\ 
2007 Oct 18    & NARVAL   & 392.33125--392.41936 & 19:50:58--21:57:51 & $3\times4\times900$ & 340--390 & 3.3 &  0.9$\pm$82.9\\ 
2007 Oct 19    & NARVAL   & 393.36308--393.45120 & 20:36:51--22:43:45 & $3\times4\times900$ & 580--620 & 2.1 & -66.6$\pm$50.4\\ 
2007 Oct 20    & NARVAL   & 394.34587--394.43399 & 20:12:07--22:19:01 & $3\times4\times900$ & 560--600 & 2.1 & -99.6$\pm$50.8\\ 
2007 Oct 21    & NARVAL   & 395.37385--395.46197 & 20:52:28--22:59:21 & $3\times4\times900$ & 560--590 & 2.1 & -215.1$\pm$52.0\\ 
2007 Oct 23    & NARVAL   & 397.35103--397.43915 & 20:19:43--22:26:37 & $3\times4\times900$ & 540--590 & 2.1 & -70.6$\pm$52.7\\ 
2007 Oct 24    & NARVAL   & 398.35262--398.44076 & 20:22:03--22:28:59 & $3\times4\times900$ & 550--600 & 2.2 & -25.6$\pm$52.9\\ 
\hline 
2008 Oct 14    & NARVAL   & 754.30940--754.48559 & 19:19:21--23:33:04 & $5\times4\times900$ & 530--610 & 1.6 & -81.0$\pm$39.2\\ 
2008 Oct 16    & ESPaDOnS & 755.85707--755.88927 & 08:28:03--09:14:25 & $2\times4\times650$ & 540--580 & 2.7 & -115.4$\pm$69.5\\ 
2008 Oct 23    & NARVAL   & 763.29085--763.46708 & 18:53:06--23:06:53 & $5\times4\times900$ & 460--550 & 1.9 & -137.0$\pm$48.0\\ 
2008 Oct 24    & NARVAL   & 764.27121--764.44746 & 18:24:52--22:38:41 & $5\times4\times900$ & 600--650 & 1.5 & -106.9$\pm$37.9\\ 
2008 Oct 25    & NARVAL   & 765.28297--765.45922 & 18:41:52--22:55:41 & $5\times4\times900$ & 420--520 & 2.0 & -56.4$\pm$51.7\\ 
2008 Oct 26    & NARVAL   & 766.26993--766.44619 & 18:23:09--22:36:59 & $5\times4\times900$ & 420--530 & 1.9 & -66.5$\pm$50.3\\ 
\hline 
2009 Jul 05    & ESPaDOnS & 1018.04324--1018.10593 & 13:00:46--14:31:02 & $2\times4\times1300$ & 950--1040 & 1.6 & -134.4$\pm$43.8\\ 
2009 Jul 07    & ESPaDOnS & 1020.04033--1020.10416 & 12:56:22--14:28:16 & $2\times4\times1300$ & 830--950 & 1.7 & -286.0$\pm$47.3\\ 
2009 Jul 09    & ESPaDOnS & 1022.04307--1022.10552 & 13:00:05--14:30:00 & $2\times4\times1300$ & 1070--1090 & 1.5 & -222.3$\pm$40.1\\ 
2009 Jul 13    & ESPaDOnS & 1026.03827--1026.10162 & 12:52:44--14:23:58 & $2\times4\times1300$ & 1030--1060 & 1.5 & -140.0$\pm$40.6\\ 
2009 Jul 14    & ESPaDOnS & 1027.05150--1027.11423 & 13:11:41--14:42:01 & $2\times4\times1300$ & 1100--1110 & 1.4 & -142.9$\pm$39.8\\ 
2009 Jul 21    & NARVAL   & 1033.53871--1033.63538 & 00:52:37--03:11:49 & $3\times4\times750$ & 330--520 & 2.7 & -130.7$\pm$66.4\\ 
2009 Jul 25    & NARVAL   & 1037.53156--1037.60579 & 00:41:56--02:28:49 & $3\times4\times750$ & 480--510 & 2.5 & -160.0$\pm$62.2\\ 
2009 Jul 26    & NARVAL   & 1038.57750--1038.64709 & 01:47:60--03:28:11 & $3\times4\times700$ & 330--450 & 3.1 & 121.0$\pm$82.4\\ 
2009 Jul 27    & NARVAL   & 1039.53964--1039.65875 & 00:53:23--03:44:53 & $4\times4\times806$ & 310--440 & 2.7 & -28.1$\pm$74.0\\ 
2009 Jul 28    & NARVAL   & 1040.53824--1040.62633 & 00:51:16--02:58:06 & $3\times4\times900$ & 260--460 & 3.2 & -113.0$\pm$87.6\\ 
2009 Jul 29    & NARVAL   & 1041.53043--1041.61852 & 00:39:56--02:46:46 & $3\times4\times900$ & 510--580 & 2.2 & -27.8$\pm$55.3\\ 
2009 Jul 30    & NARVAL   & 1042.54192--1042.63003 & 00:56:23--03:03:16 & $3\times4\times900$ & 580--640 & 2.0 & -62.7$\pm$50.7\\ 
2009 Jul 31    & NARVAL   & 1043.52399--1043.64394 & 00:30:28--03:23:11 & $4\times4\times812$ & 500--550 & 2.0 & -161.6$\pm$51.3\\ 
2009 Aug 01    & NARVAL   & 1044.54174--1044.62418 & 00:55:57--02:54:39 & $3\times4\times812$ & 540--590 & 2.1 & -38.0$\pm$55.1\\ 
2009 Aug 02    & NARVAL   & 1045.53783--1045.62615 & 00:50:14--02:57:24 & $3\times4\times900$ & 390--420 & 3.1 & -99.4$\pm$80.0\\ 
2009 Aug 04    & NARVAL   & 1047.56189--1047.64999 & 01:24:42--03:31:34 & $3\times4\times900$ & 560--600 & 2.1 & -161.4$\pm$36.6\\ 
2009 Aug 05    & NARVAL   & 1048.51890--1048.60701 & 00:22:43--02:29:35 & $3\times4\times900$ & 580--600 & 2.1 & -141.9$\pm$52.7\\ 
2009 Sep 03    & ESPaDOnS & 1078.07457             & 13:41:03           & $4\times1300$       & 940 & 2.2 & -109.7$\pm$55.8\\ 
2009 Sep 08    & ESPaDOnS & 1082.95060--1083.01300 & 10:42:24--12:12:14 & $2\times4\times1300$ & 1030--1040 & 1.5 & -156.1$\pm$38.2\\ 
2009 Sep 25    & ESPaDOnS & 1099.88204--1099.94611 & 09:03:29--10:35:43 & $2\times4\times1300$ & 990--1010 & 1.5 & -208.5$\pm$40.3\\ 
2009 Oct 01    & ESPaDOnS & 1105.86247--1105.92525 & 08:35:21--10:05:45 & $2\times4\times1300$ & 880--960 & 1.7 & -140.4$\pm$45.2\\ 
2009 Oct 05    & ESPaDOnS & 1109.96249--1110.02567 & 10:59:28--12:30:27 & $2\times4\times1300$ & 140--600 & 3.2 & -273.8$\pm$84.1\\
2009 Oct 10    & ESPaDOnS & 1114.87331--1114.93701 & 08:51:12--10:22:56 & $2\times4\times1300$ & 1060--1100 & 1.6 & -239.7$\pm$41.2\\
\hline
\end{tabular}
\label{tab:log}
\end{table*}

\section{Magnetic detection}
\label{s_mag}

Least-Squares Deconvolution (LSD; \citealt{Donati97}) was applied to
all observations.  The line list was constructed manually to include
the few moderate to strong absorption lines that are no more than
moderately affected by the wind.  All lines showing strong emission
from the wind and/or circumstellar environment at the time of our
observations (e.g., H$\alpha$) were excluded from the list. The
remaining very weak contamination is not circularly polarised. Hence
it only slightly affects the calculation of the longitudinal field but
not the Stokes V profile and thus the field detection. The C~{\sc iv}
lines at 5801.3 and 5812.0 \AA\ are used as reference photospheric
lines from which we obtain the average radial velocity of HD~108
(about -62~\kms); about a handful of unblended absorption lines that
are not blueshifted with respect to the reference frame by more than
30~\kms\ are also included in the list.  We ended up with a list of 16
lines, whose characteristics are summarized in Table~\ref{tab:lines}.
All LSD profiles were produced on a spectral grid with a velocity bin
of 7.2~\kms, providing reasonable sampling given the significant
macroturbulent broadening detected on HD~108 (see
Sect.~\ref{sec:rot}).

\begin{table}
\caption[]{Lines used for Least-Squares Deconvolution.  The
line depths (column 3) were directly measured from our spectra while
the Land\'e factors (column 4) were derived assuming LS coupling. Bold face indicates lines with the weakest wind contamination } 
\begin{tabular}{clcc}
\hline
Wavelength & Element & Depth & Land\'e   \\
(\AA)       &         & ($I_{\rm c}$) & factor      \\
\hline
 4026.198 & \hei   & 0.33 &   1.1 \\     
 4199.839 & \heii  & 0.15 &   1.0 \\    
 4379.201 & \niii  & 0.06 &   1.1 \\   
 4387.929 & \hei   & 0.08 &   1.0 \\   
 4471.483 & \hei   & 0.41 &   1.1 \\    
 4510.963 & \niii  & 0.06 &   1.1 \\     
 4514.854 & \niii  & 0.08 &   1.2 \\      
 \textbf{4541.593} & \textbf{\heii}  & \textbf{0.20} &   \textbf{1.0} \\     
 4713.139 & \hei   & 0.13 &   1.5 \\      
 4921.931 & \hei   & 0.14 &   1.0 \\     
 5015.678 & \hei   & 0.11 &   1.0 \\     
 \textbf{5411.516} & \textbf{\heii}  & \textbf{0.25} &   \textbf{1.0} \\     
 \textbf{5592.252} & \textbf{\oiii}  & \textbf{0.40} &   \textbf{1.0} \\     
 \textbf{5801.313} & \textbf{\civ}   & \textbf{0.20} &   \textbf{1.2} \\    
 \textbf{5811.970} & \textbf{\civ}   & \textbf{0.15} &   \textbf{1.3} \\    
 7065.176 & \hei   & 0.22 &   1.5 \\     
\hline
\end{tabular}
\label{tab:lines}
\end{table}

Using this line list, we obtained, for all collected spectra, mean
circular polarization (LSD Stokes $V$), mean polarization check (LSD
$N$) and mean unpolarized (LSD Stokes $I$) profiles (corresponding to
a weighted -- relative to \sn$^2$, line depth and
  wavelength -- average line of central wavelength 5500 \AA\ and
Land\'e factor 1.2).  Averaging together all LSD profiles recorded on
each night of observation (with weights proportional to the inverse
variance of each profile) yields relative noise levels ranging from
1.4 to 3.3 (in units of \ptt) in $V$ and $N$ profiles. In general, no
significant magnetic signature was detected in individual spectra. In
a few exceptional cases, however, nightly averages of spectra showed
marginal evidence for the presence of magnetic fields at the surface
of HD~108 (see column 8 of Table \ref{tab:log}).

Using a line list containing the five lines with weakest
  interstellar contamination and smallest blueshifts with respect to
  the stellar rest velocity (indicated in bold face in Table
  \ref{tab:lines}), the detected Zeeman signature is compatible
  (within noise level) with that derived using the full line list,
  confirming that the detected signature is due to magnetic fields at
  the surface of the star.

The detected Zeeman signature (featuring a peak-to-peak amplitude of 
 0.04\%) cannot be attributed to the small crosstalk between circular and 
 linear polarisations that ESPaDOnS and NARVAL used to suffer (smaller 
 than 1\% on ESPaDOnS during our 2009 run, smaller than 4\% for NARVAL 
 and for ESPaDOnS at all other epochs) as this would imply that 
 photospheric lines of HD~108 exhibit roughly antisymmetric linear 
 polarisation signatures with peak-to-peak amplitudes of at least 1-4\% 
 - a highly unlikely possibility.  It would also imply that the Zeeman  
 signature detected with ESPaDOnS in 2009 is smaller (by at least a factor 
 of 4) than those measured with NARVAL and ESPaDOnS at all other epochs, 
 reflecting the instrumental fix implemented on ESPaDOnS in early 2009;  
 since this is not the case, we can safely claim that the detected Zeeman 
 signature is not caused by a crosstalk problem.

\begin{figure}
\center{\includegraphics[scale=0.35,angle=-90]{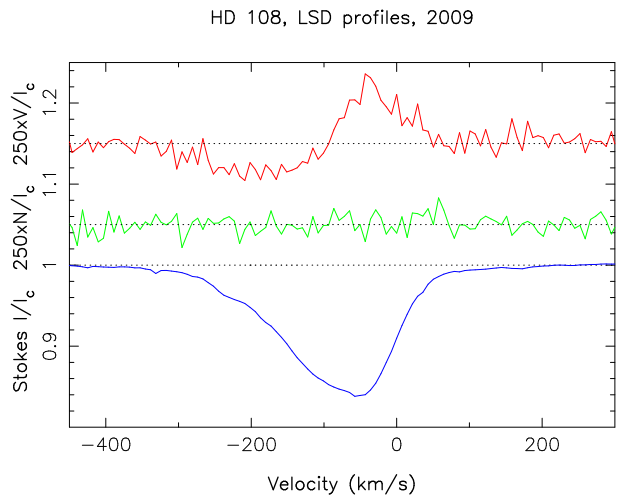}}
\caption[]{LSD circularly-polarised (Stokes $V$), polarisation check ($N$) and 
unpolarised (Stokes $I$) profiles of HD~108 (top/red, middle/green, 
bottom/blue curves respectively) averaged over all NARVAL and ESPaDOnS 
data collected in 2009.  LSD $V$ and $N$ profiles are expanded by a factor 
of 250 and shifted upwards (by 1.15 and 1.05) for display purposes.}
\label{fig:lsd}
\end{figure}

For each of the main observing seasons (2007, 2008 and 2009), we
constructed average LSD signatures and corresponding variances.  In
2009, the mean Stokes $V$ signature is clearly detected with a reduced
chi-squared \chisqr\ within the line profile of about 13\footnote{The
  reduced chi-squared estimator quantifies how significantly the
  observed Zeeman signature departs from a null profile
  (i.e. $V$=0).)} (see Fig.~\ref{fig:lsd}); in 2008 and 2007, the
corresponding \chisqr\ values (2.5 and 1.5 respectively) are smaller
but large enough to claim a definite detection in 2008 and a marginal
one in 2007 (with a 98\% confidence level).  No signal is detected in
the $N$ (polarization check) profile, demonstrating that the signal we
detect is truly attributable to circular polarization.  The variance
profiles show no evidence for variability of the Zeeman signatures
throughout each season, even in 2009 where the signal to noise ratio
is highest.  We therefore conclude that the Zeeman signature, if
varying, is changing very slowly on typical timescales of years rather
than months or weeks.  The longitudinal fields we measure in 2007,
2008 and 2009 \citep[using the first order moment method of][-- inspired by Semel 1967 -- and
  integrating through a velocity range of --300 to 120~\kms]{Donati97}
are respectively equal to $-90\pm20$, $-100\pm20$ and $-150\pm10$~G
(1$\sigma$ error bars), suggesting a slow rise in field strength on
timescales of years.

The detection of a magnetic field on HD~108 as well as the lower limit
of months to years on the timescale on which the Zeeman signatures are
varying both strengthen the aforementioned analogy with \tori\ and
HD~191612. For those stars, the longitudinal fields and line emissions (and
photometric brightness for HD~191612) are all modulated with the same
period (the rotation period).  We thus speculate that the longitudinal
field of HD~108 tightly correlates with line emission and photospheric
brightness, weakest at minimum emission and minimum brightness and
strongest at maximum emission and maximum brightness (as in
HD~191612). This is in agreement with the slow increase in both longitudinal
field and Balmer line emission that we observe from 2007 to 2009.

If we assume that the magnetic field at the surface of HD~108 is 
grossly dipolar \citep[e.g., like that of \tori\ and 
HD~191612,][]{donati02, donati06}, it implies that 
the polar field at the surface of the star \citep[roughly 4 times 
larger than the maximum longitudinal field, e.g.,][]{donati02} is 
at least 0.5--0.6~kG. This estimate is based on the classical relation between longitudinal and polar field of \citet{preston67} \citep[see also][]{dl09}. If we further take into account that HD~108 
is still at least 10~yr away from longitudinal field maximum 
\citep[e.g. given the observed photometric variations,][]{barannikov07}, 
we conclude that the polar field of HD~108 is likely stronger than 1~kG 
and potentially as strong as (or even stronger than) that of HD~191612.

\section{STELLAR AND WIND PHYSICAL PROPERTIES}
\label{s_models}

\citet{naze08} conducted a study of the ``high'' optical state of
HD~108. Here, we present a new analysis based on our NARVAL spectra
collected in 2008 and 2009, as well as {\it FUSE} data obtained in
2007 and 2008 and retrieved from the MAST archive. These data
correspond to the ``low'' state of HD108.  The spectral analysis was
performed using the atmosphere code CMFGEN \citep{hm98}.   

An exhaustive description of the atmosphere models can be found in \citet{hm98}. In practice, an iterative scheme is developed. At each step, the radiative transfer equation is solved in spherical geometry. With the resulting specific intensity, the rate equations yield non--LTE level populations which in turn are used to compute opacities. These opacities are subsequently injected in the radiative transfer equation in the next iteration step. The temperature structure results from the radiative equilibrium equation. Line--blanketing is included by means of the super--level approach. The atmosphere velocity structure is given as input and results from the connection of a pseudo--hydrostatic structure \citep[taken from TLUSTY models,][]{lh03} with a so--called $\beta$--velocity law representing the wind structure \citep[e.g.][]{lc99}. The density structure directly follows from mass conservation. 
All models take into account clumping, and Auger
ionization by X-rays \citep[using the X--ray luminosity
  of][]{naze04}.  We adopt a solar chemical composition
\citep{grevesse07} except for helium and nitrogen (see below). Once the model atmosphere is obtained, a formal solution of the radiative transfer equation is performed with to produce the final synthetic spectra.

The derived stellar and wind properties of HD~108 are gathered in Table
\ref{params}. The results of \citet{naze08} are also given for
  comparison.

\subsection{Photospheric parameters}      

\begin{figure}
\center
\includegraphics[width=9cm,height=11cm]{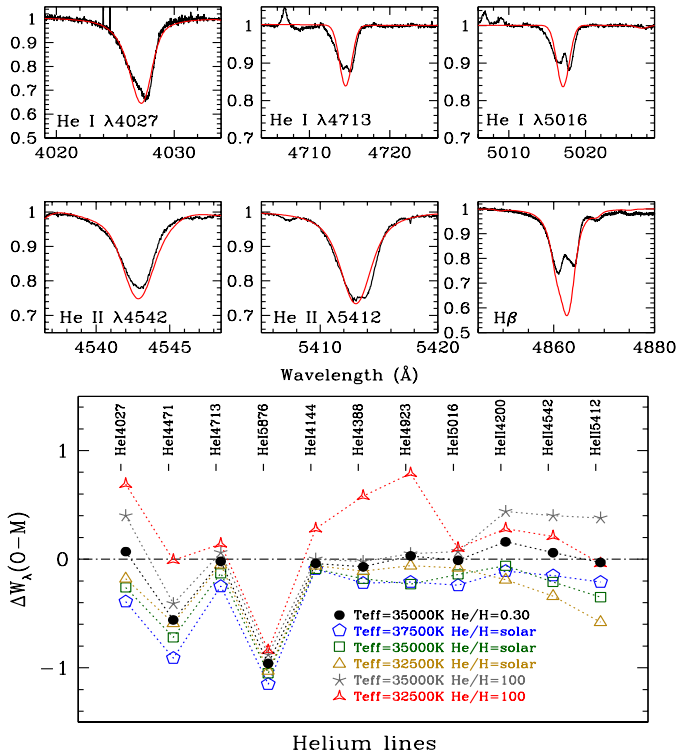}
\caption{Determination of the effective temperature and surface gravity of
  HD~108. Upper panels: final model from CMFGEN (red/grey) and observed spectrum (black). Lower panel:
observed minus synthetic equivalent widths of several Helium lines ($\Delta W_\lambda$ (O-M) $= W^{obs}_\lambda - W^{model}_\lambda$).
A perfect agreement is illustrated by the dash-dotted line.}
\label{figteff}
\end{figure}

The effective temperature \teff\ is usually derived from the relative
strength of He~{\sc i} to He~{\sc ii} lines in O stars. In the case of
HD~108 however, this procedure turned out to be considerably more
complex. First, some He~{\sc i} lines present a central emission
component the origin of which is not known (e.g. He~{\sc
  i}~$\lambda$~4144, He~{\sc i}~$\lambda$~4388, and He~{\sc
  i}~$\lambda$~4920). Second, most lines have highly asymmetric
profiles, their blue wing rising towards the continuum less steeply
than their red wings (see e.g. Fig.\ \ref{fig:lsd}). Third, He~{\sc
  i}~$\lambda$~5876 and, more worrisome, the normally well--behaved
He~{\sc i}~$\lambda$~4471 line show unusually strong absorption in the
2007--2009 spectra. In fact, these two lines are the strongest
features of the entire optical spectrum.  Fig.\ \ref{figteff}
illustrates the problems encountered when fitting the He~{\sc i}
lines. It shows the difference between the observed He {\sc i}
equivalent widths and those of various models. Clearly, He~{\sc
  i}~$\lambda$~4471 and He~{\sc i}~$\lambda$~5876 are always too weak
in our models, because both the line depth and equivalent width are weaker than observed. The situation is improved only for very high He/H
ratios ($>$ few tens), but such values are unrealistic and provide a
worse fit to other He lines.  We thus decided to leave these lines
aside and focused on the remaining He {\sc i} and He~{\sc ii}
lines. The best fit was achieved for \teff\ = 33000--37000 K and a
He/H ratio of 0.3 (by number). It is shown in Fig.\ \ref{figteff} with
\teff\ = 35000 K.  We also needed a larger than solar N abundance to
fit the N~{\sc iii} lines in the interval 4510-4540\AA, namely, N/H =
$3.6 \times 10^{-4}$ ($\sim$6x solar by number). A discussion of the abundance pattern of magnetic O stars in relation with evolutionary models will be presented in a subsequent publication (Escolano et al.\ 2010, in prep).

\begin{figure}
\center
\includegraphics[width=9cm]{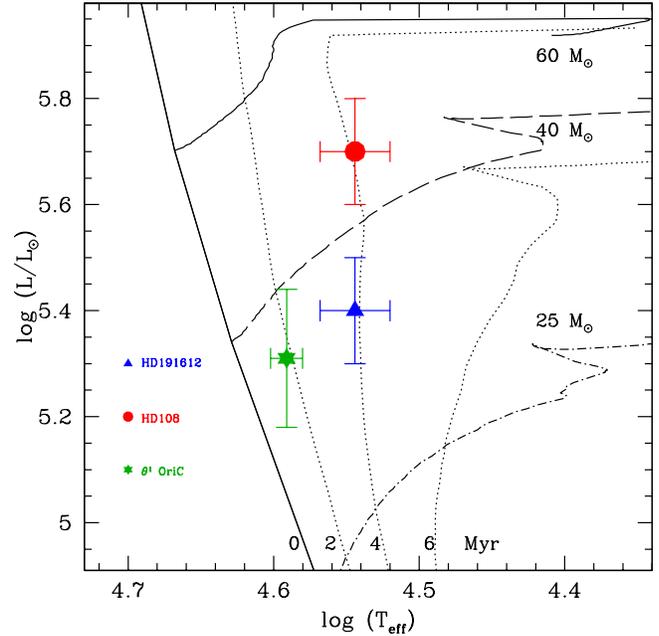}
\caption{HR diagram with the position of HD~108 (red circle), HD~191612 (blue triangle) and \tori\ (green asterisk) indicated. Evolutionary tracks include stellar rotation and are from \citet{mm03}. Isochrones are shown as dotted lines. Stellar parameters for HD~191612 and \tori\ are from \citet{walborn03} and \citet{sergio06} respectively. }
\label{hrd}
\end{figure}

HD~108 is a member of the OB association Cas OB5. The luminosity
$L_\star$ was determined by assuming a distance of 2.51$\pm$0.15 kpc
\citep{humph78} and fitting the spectral energy distribution using 
{\it IUE} data and UBVJHK photometry. The uncertainty on \lL\ is 
of about 0.1 dex and is entirely dominated by the uncertainty on 
the distance.
The surface gravity (log $g$) was obtained from the fit to the
H$\beta$ and H$\gamma$ line wings. A value of 3.5 \citep[typical of
 late type giants/supergiants; see][]{msh05} is derived, with an uncertainty of about
0.2 dex.
The position of HD~108 in the HR diagram is shown in Fig.\ \ref{hrd}. HD~191612 and \tori\ are also included for comparison purpose. HD~108 is about 4 Myr old just as HD~191612, reinforcing the analogy between both stars. Given its slightly higer mass, HD~108 might be a little more evolved than HD~191612. HD~108 is also older than \tori, consistent with the scenario according to which \tori\ represents a precursor of HD~191612 \citep{donati06} and thus HD~108.

The comparison of the results of the present study to the one by
\citet{naze08} shows that the derived stellar parameters are in rather
good agreement in spite of the difficulties encountered in the present
analysis. Only the luminosity is marginally higher in the present study
\citep[but the method we used for its determination is more accurate
 than in][]{naze08}. This overall similarity demonstrates that the
variability altering the shape of the optical lines is probably not
due to changes in the star's stellar parameters. Instead, they might
be related to the wind or immediate environment of the star.

\begin{table}
\caption{Summary of stellar and wind properties of HD~108.}
\begin{tabular}{lll}
                         &   present work      &    Naz\'e et al.\ 2008 \\
\hline
T $_{eff}$ (K)            & 35 000 $\pm$ 2000   &  37000 $\pm$ 2000 \\
log $g$ (cgs)            & 3.50 $\pm$ 0.20     &  3.75 $\pm$ 0.10    \\
log L$_\star$/L$_\odot$   & 5.70  $\pm$ 0.10    &  5.40 $\pm$ 0.10   \\
R$_{\star}$/R$_\odot$     & 19.2 $^{+3.3}_{-2.8}$  & 12.3 $\pm$ 2.1 \\
M$_{\star}$/M$_{\odot}$    & 43 $^{+32}_{-18}$     &  $\sim$ 35   \\
v$sini$ (km s$^{-1}$)    &  $<$ 50              & $\sim$ 40 \\
log $\dot{M}$           & -7.0$^{+0.20}_{-0.40}$ & -6.5/-7.0  \\
v$_\infty$ (km s$^{-1}$)  & 2000 $\pm$ 300      & 2000  \\
v$_{cl}$ (clumping; km s$^{-1}$) & 30           &  30  \\
$f$ (clumping)          & 0.01                 & 0.01  \\
\hline
E(B-V)                  & 0.47                 & -- \\
log $L_X/L_{BOL}$        & -6.2                 & -- \\
He/H (by number)        & $3.0 \times 10^{-1}$  & 0.1$^{*}$ \\
N/H (by number)         & $3.6 \times 10^{-4}$  & -- \\
\hline
\end{tabular}\\
{$*$ adopted} \\
\label{params}
\end{table}

\subsection{Wind parameters}
\label{s_wind}

We relied on strong EUV and FUV lines to constrain the mass-loss rate
($\dot{M}$) and the terminal velocity (v$_\infty$). We avoided the use
of wind sensitive optical lines (H$\alpha$ and He~{\sc
  ii}~$\lambda$~4686) since they show a high degree of short-term
variability and can be contaminated by non-stellar emission.

\begin{figure}
\center
\includegraphics[width=9.5cm,height=9cm]{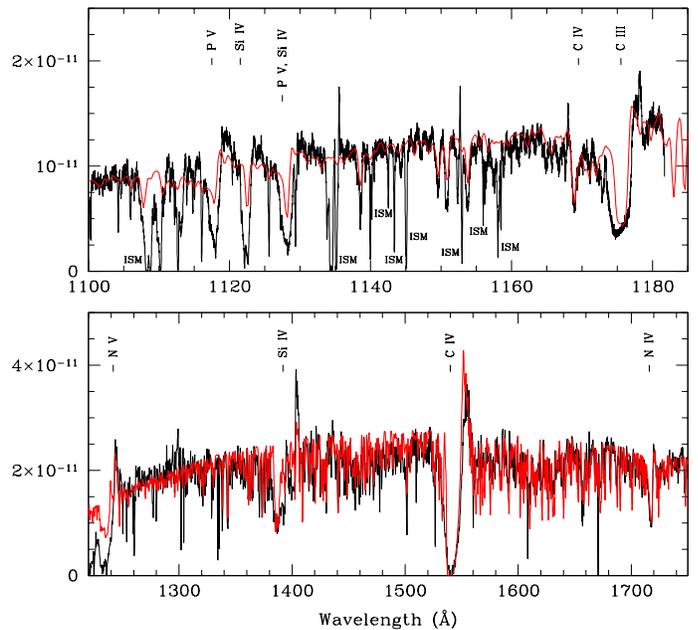}
\caption{UV and far-UV observed spectra (black) and our final CMFGEN model (red). Top: {\it FUSE} LiF2A spectrum. Bottom: {\it IUE} spectrum. The region below 1200 \AA\ common to the FUSE spectrum is very noisy and is not shown. The flux is in units of erg s$^{-1}$ cm$^{-2}$ \AA$^{-1}$.}
\label{uvfit}
\end{figure}

We used only the {\it FUSE} LiF2A channel of the {\it FUSE} spectra in
our analysis since wavelengths shorter than 1100\AA\, are heavily
contaminated by the ISM. Our best fit is shown in Fig.\ \ref{uvfit}
(upper panel) and the corresponding parameters are listed in Table
\ref{params}. The synthetic P~{\sc v} $\lambda\lambda$ 1118,28 and
Si~{\sc iv} $\lambda\lambda$1122,28 lines are somewhat weaker and
narrower than observed. By decreasing the mass-loss rate we tend to
decrease this discrepancy, but a worse fit is achieved to the C~{\sc
  iii} $\lambda$1176 and C~{\sc iv} $\lambda$1169 transitions.

Unfortunately, no FUV spectra (1200 to 2000 \AA) exist for the ``low''
optical state. The only {\it IUE} observations cover the period
1978-1989 and are quite similar, showing no
  signs of significant variability. They presumably correspond to
HD~108's "high" optical state \citep[the Balmer lines being in
  emission at least between 1987 and 1989,
  see][]{underhill94,naze01}. However, we found out that the set of
parameters derived from the {\it FUSE} spectra lead to a good fit of
the IUE data (see lower panel of Fig.\ \ref{uvfit}). We thus conclude
that the global wind properties of HD~108 do not vary significantly
between the low and high state. In particular, the mass loss rate does
not differ by more than a factor of two (corresponding to the
uncertainty on our determination). We note that the wind parameters
are also consistent with those derived by \citet{naze08} who used the
same {\it IUE} data but different optical spectra.

Given that the wind properties do not seem to depend on the
  star's state (high or low), we used N~{\sc iv} $\lambda$1718 to
  constrain the clumping factor \citep[see a complete description
    in][]{jc05}. A small $f$ value (0.01) corresponding to a large
  degree of wind inhomogeneities was necessary to correctly reproduce
  the blue wing extension.

\subsection{Rotational velocity and macroturbulence}
\label{sec:rot}

We used the Fourier transform method to constrain the projected
rotational velocity \citep{gray92}. To avoid any problem caused by
wind and/or circumstellar emission, we selected photospheric lines
showing the least degree of contamination. It turned out that no line
showed a pure symmetric profile. Even for the best candidate features,
the blue wing clearly showed a wider extension than the red wing,
indicating some kind of contamination, most likely due to the
wind. Consequently, we decided to work on an artificial, symmetric
line profile created from the red wing of C~{\sc iv} $\lambda$ 5812,
which we assume better reflects the true photospheric properties. We
used the average of the July--August 2009 NARVAL spectra as input. No
minimum was seen in the Fourier transform above the noise level,
indicating that 1) the maximum \vsini\ compatible with our artificial
profile is $\sim$ 50 \kms\ and 2) that macroturbulence is an important
broadening mechanism. Tests run on N~{\sc iv} $\lambda$ 4057 and
O~{\sc iii} $\lambda$ 5592 gave similar results. Note that the origin of macroturbulence in O stars is widely unknown, although \citet{aerts09} recently claimed that non-radial pulsations could be an important ingredient.

\begin{figure*}
\begin{center}
\begin{minipage}[b]{0.32\linewidth} 
\centering
\includegraphics[width=6cm]{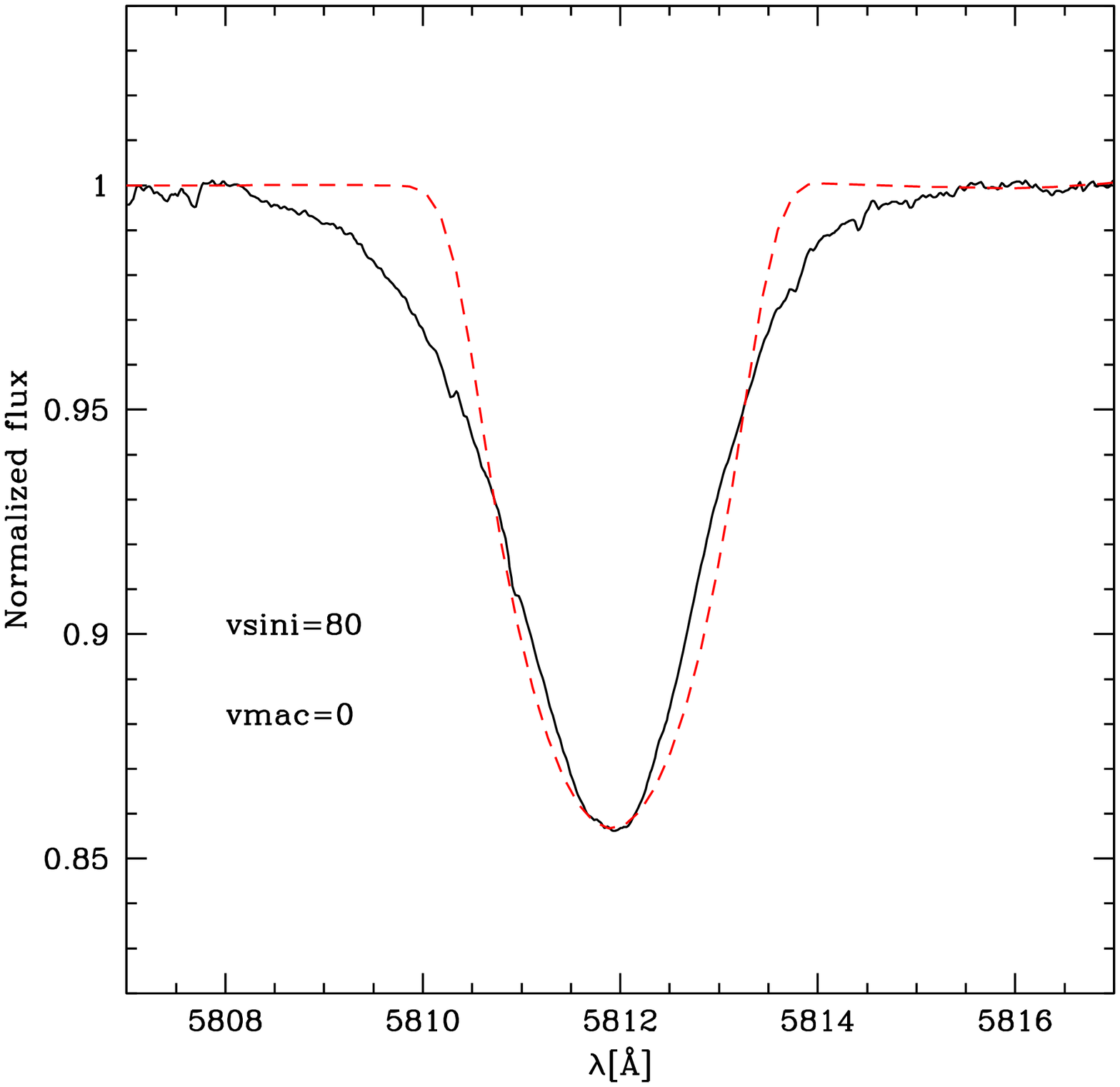}
\end{minipage}
\hspace{0.05cm} 
\begin{minipage}[b]{0.32\linewidth}
\centering
\includegraphics[width=6cm]{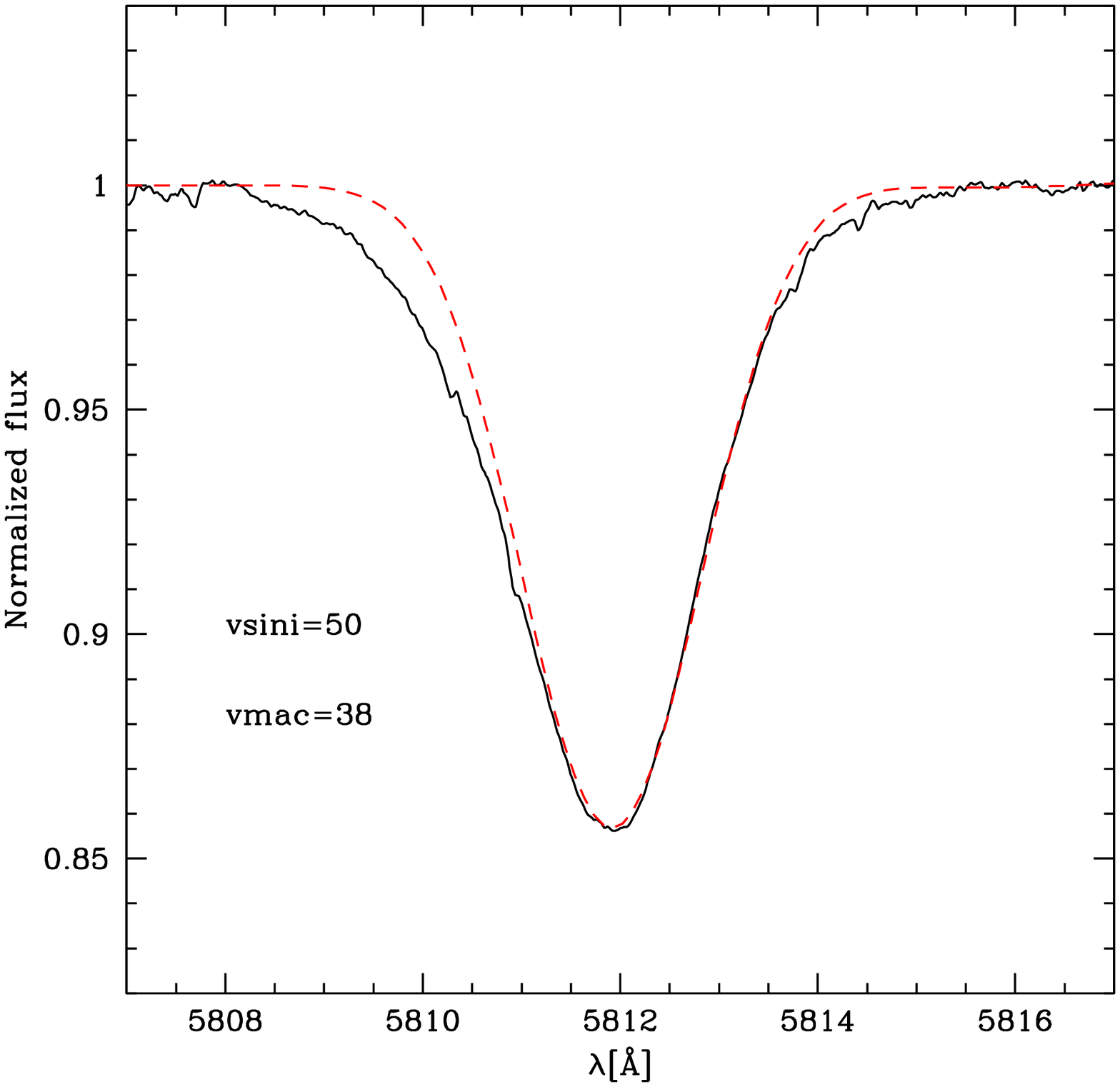}
\end{minipage}
\hspace{0.05cm} 
\begin{minipage}[b]{0.32\linewidth}
\centering
\includegraphics[width=6cm]{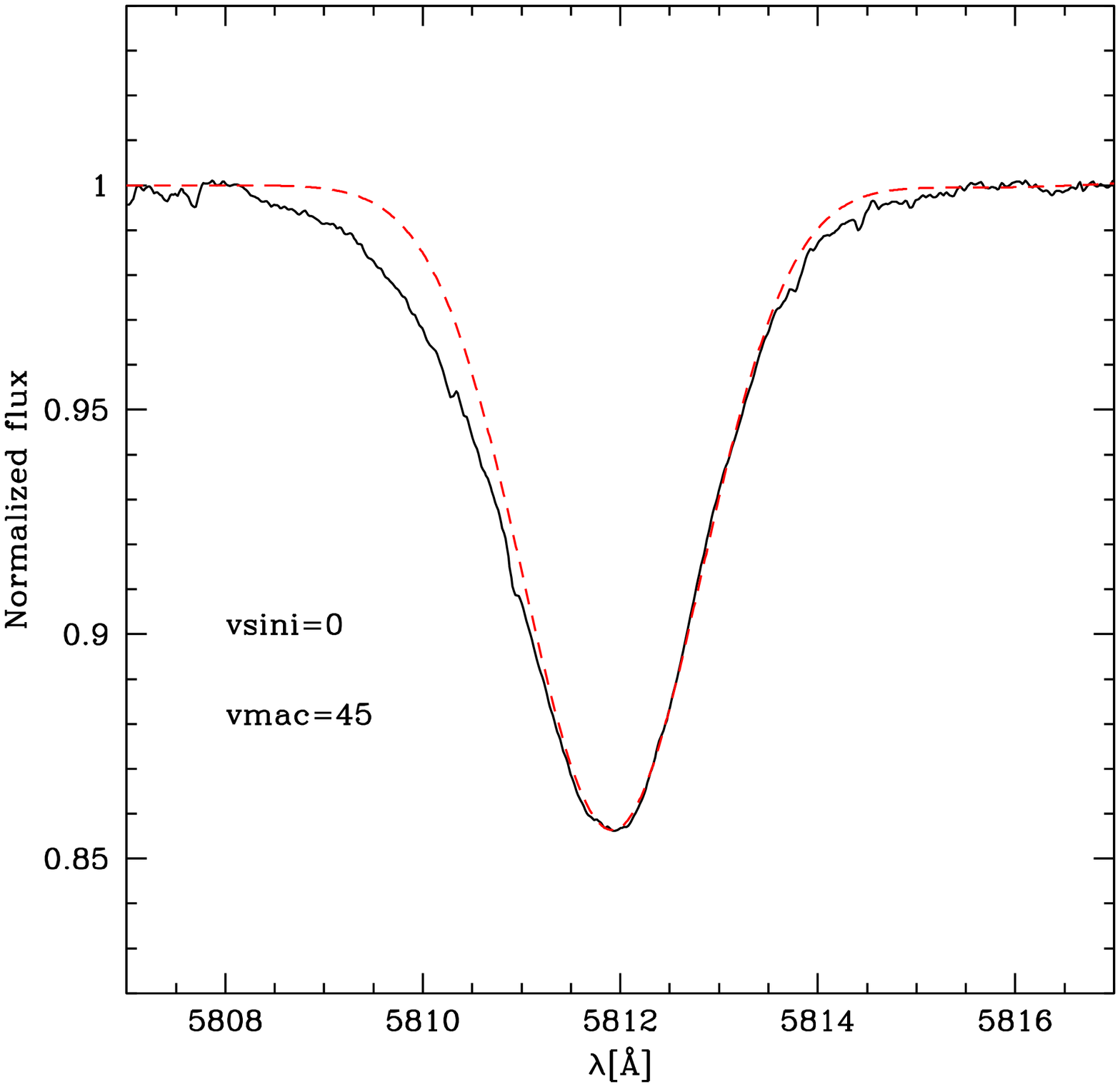}
\end{minipage}
\caption{Comparison of the C~{\sc iv} $\lambda$ 5812 line with synthetic profiles with the following \vsini/v$_{\rm mac}$ values (in \kms): 80/0 (left), 50,38 (middle), 0/45 (right). The black solid line is the observed profile and the red dashed line the synthetic profile. While \vsini\ larger than 50 \kms\ is excluded, any value below this limit gives a satisfactory fit to the red wing provided a significant amount of macroturbulence is included.}\label{vsinimac}
\end{center}
\end{figure*}

To further investigate the rotational velocity and to quantify the
amount of macroturbulence, we used a synthetic line profile (taken
from the TLUSTY OSTAR2002 grid of models).  We artificially scaled it
in order to match the equivalent width of the observed C~{\sc iv}
$\lambda$ 5812 line. Since macroturbulence is the main broadening
mechanism (see below) this is justified, the intrinsic profile being
unimportant. We subsequently convolved this synthetic profile with 1)
a rotational profile (characterized by \vsini) and 2) a pure Gaussian
profile (characterized by its FWHM and the corresponding
macroturbulence velocity v$_{\rm mac}$). The latter mimics the effect
of isotropic turbulence. Several (\vsini\ / FWHM) combinations were
used and compared to the artificial profile described above. We found
that rotational velocities larger than about 50 \kms\ are indeed
excluded and that a macroturbulence larger than $\sim$ 30 \kms\ was
necessary to reproduce the wings slope and extension. Different
combinations of (\vsini\ / v$_{\rm mac}$) gave fits of similar
quality. Fig. \ref{vsinimac} shows three examples with \vsini\ = 0, 50
and 80 \kms\ and the corresponding v$_{\rm mac}$ = 45, 38, 0
\kms. HD~108 is therefore very similar to HD191612 \citep{howarth07}
which also features macroturbulence--dominated line profiles.

\section{DISCUSSION}
\label{s_disc}

The analogy between HD~108 and \tori/HD~191612 suspected in Sect.\ \ref{s_intro} seems to be confirmed by the present study. In particular, the similarity with HD~191612 is striking. Both stars show the same long-term spectroscopic variability, with He~{\sc i} and Balmer lines (other than \hal) changing from P--Cygni to absorption--dominated profiles \citep[see Table 6 of][]{naze01}. \citet{howarth07} reported a strong correlation between \hal\ emission and photometry, HD~191612 being brighter in the ``high'' state. \citet{barannikov07} provides evidence for the same behaviour of HD~108: photometry over the last 15 years reveals a fading of HD~108 in parallel to a decrease of the emission component of He~{\sc i} and Balmer lines (see also Naz\'e et al.\ in prep.). 

A notable difference is the variability timescale. While a period of 538 days is identified for HD~191612, a value of several decades is preferred for HD~108. This rough estimate relies on two facts. First, the photometric survey of \citet{barannikov07} discussed above points to a timescale of at least 15 years (and of at least 30 years assuming roughly sinusoidal photometric variations). Second, a transition between absorption/P--Cygni of several line profiles was observed at least twice during the last century, pointing to a periodicity of about 50--60 years \citep[see Table 6 of][]{naze01}. If, as suggested by \citet{donati06} for HD~191612, the variability timescale corresponds to the rotation period, then HD~108 would be an extremely slow rotator with an equatorial velocity $\lesssim$0.1 \kms\ \citep[much lower than predicted by evolutionary models, see Fig.\ 1 of][]{mm03}. Our upper limit on the projected rotational velocity is fully consistent with that scenario. One might thus suspect that HD~108 experienced a strong magnetic braking during its evolution.

Using the stellar and wind parameters derived in Sect.\ \ref{s_models} and the minimum polar field discussed in Sect.\ \ref{s_mag}, we obtain a confinement parameter $\eta_{\star} = \frac{B^{2} R^{2}}{\dot{M} v_{\infty}} \geq 100$. Such a value implies that the wind of HD~108 is magnetically confined \citep{ud02}. If the polar field is as large as that of HD~191612 (1.5~kG, see Sect.\ \ref{s_mag}), this would imply $\eta_{\star}$ $\sim$ 800 and thus a strong wind confinement.

According to \citet{ud09} \citep[see also discussion by][]{donati06} the typical spin-down timescale due to magnetic braking of a hot star with a strongly magnetically confined wind is $\tau_{spin} = k \frac{M}{\dot{M}} \frac{1}{\sqrt{\eta_{\star}}}$ (k is a constant with a typical value of 0.1 to 0.3). For a kG polar field this means $\tau_{spin} \sim 1-3$Myr. Given its \teff\ and luminosity, HD~108 is 3 to 5 Myr old (see Fig.\ \ref{hrd}). According to this simple estimate the rotation rate of HD~108 could have been reduced by a factor 3 to 200 just by magnetic braking (implying an initial rotational velocity of a few to 20 \kms). However, tailored simulations relying on additional observations providing a better characterization of the field strength and geometry are obviously required to quantitatively tackle the question of magnetic braking \citep{mm05,ud09}. 

\begin{figure*}
     \centering
     \subfigure[\hal]{
          \includegraphics[width=.31\textwidth]{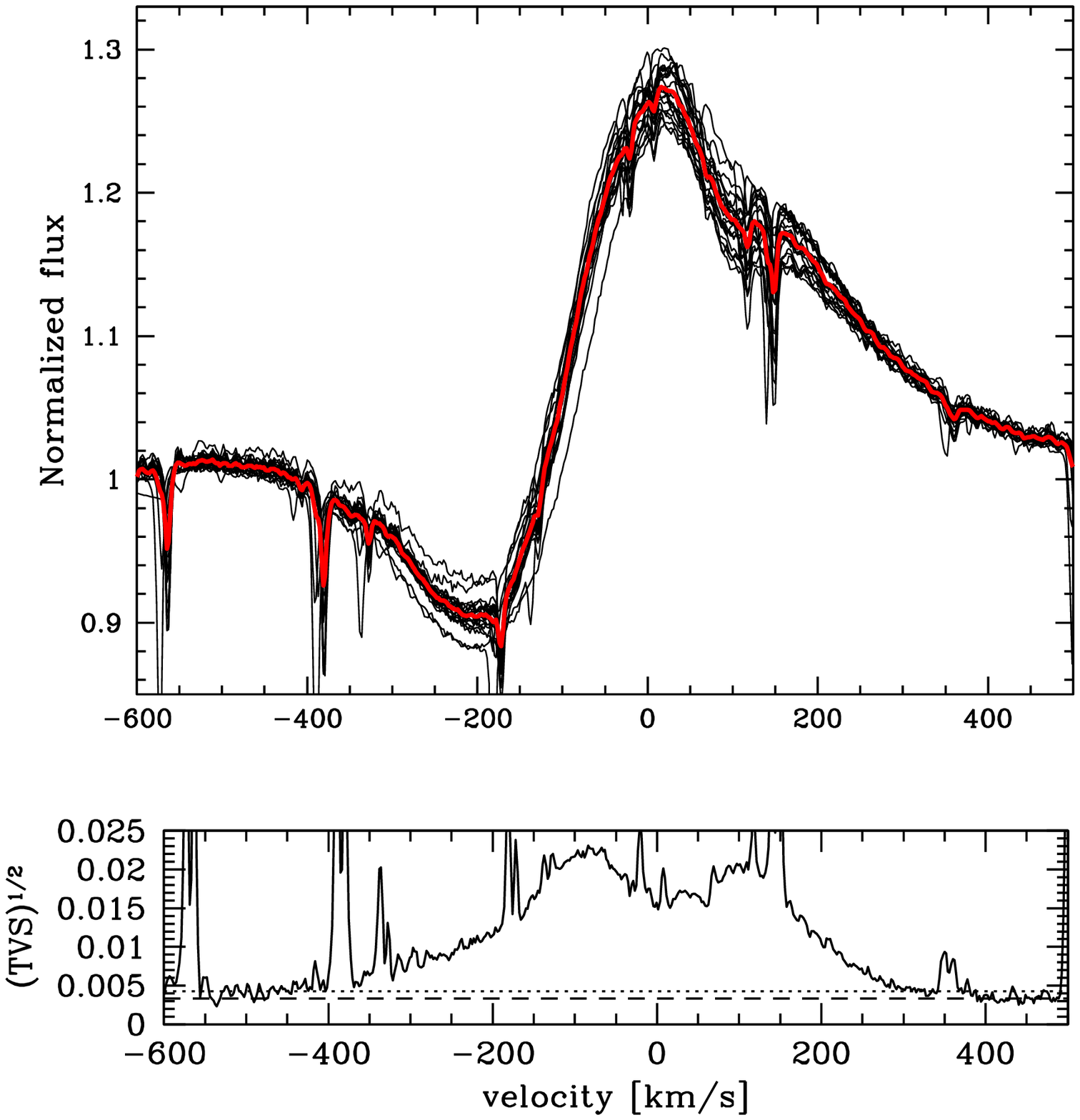}}
     \hspace{0.2cm}
     \subfigure[H${\beta}$]{
          \includegraphics[width=.31\textwidth]{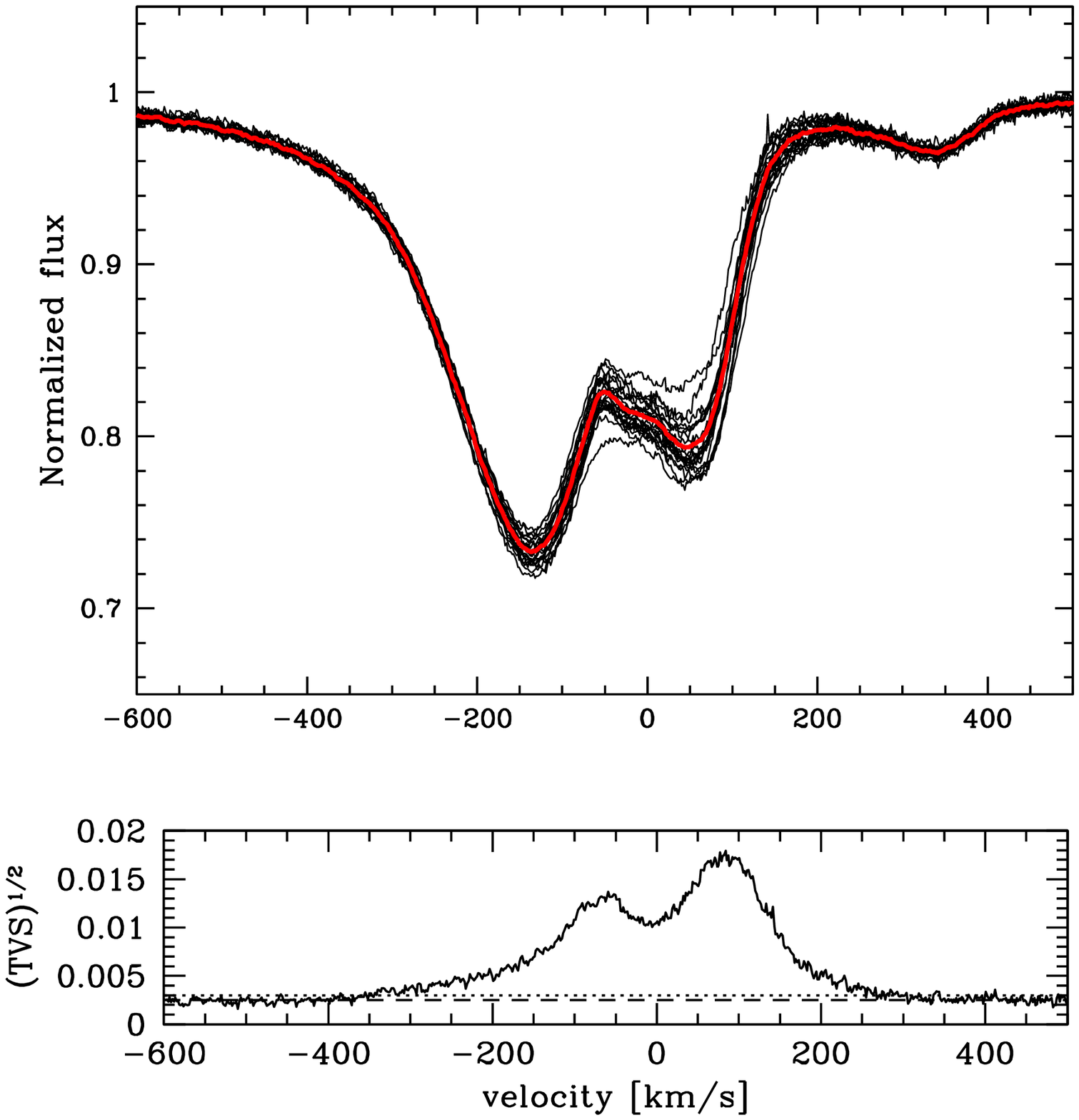}}
     \hspace{0.2cm}
     \subfigure[H${\gamma}$]{
          \includegraphics[width=.31\textwidth]{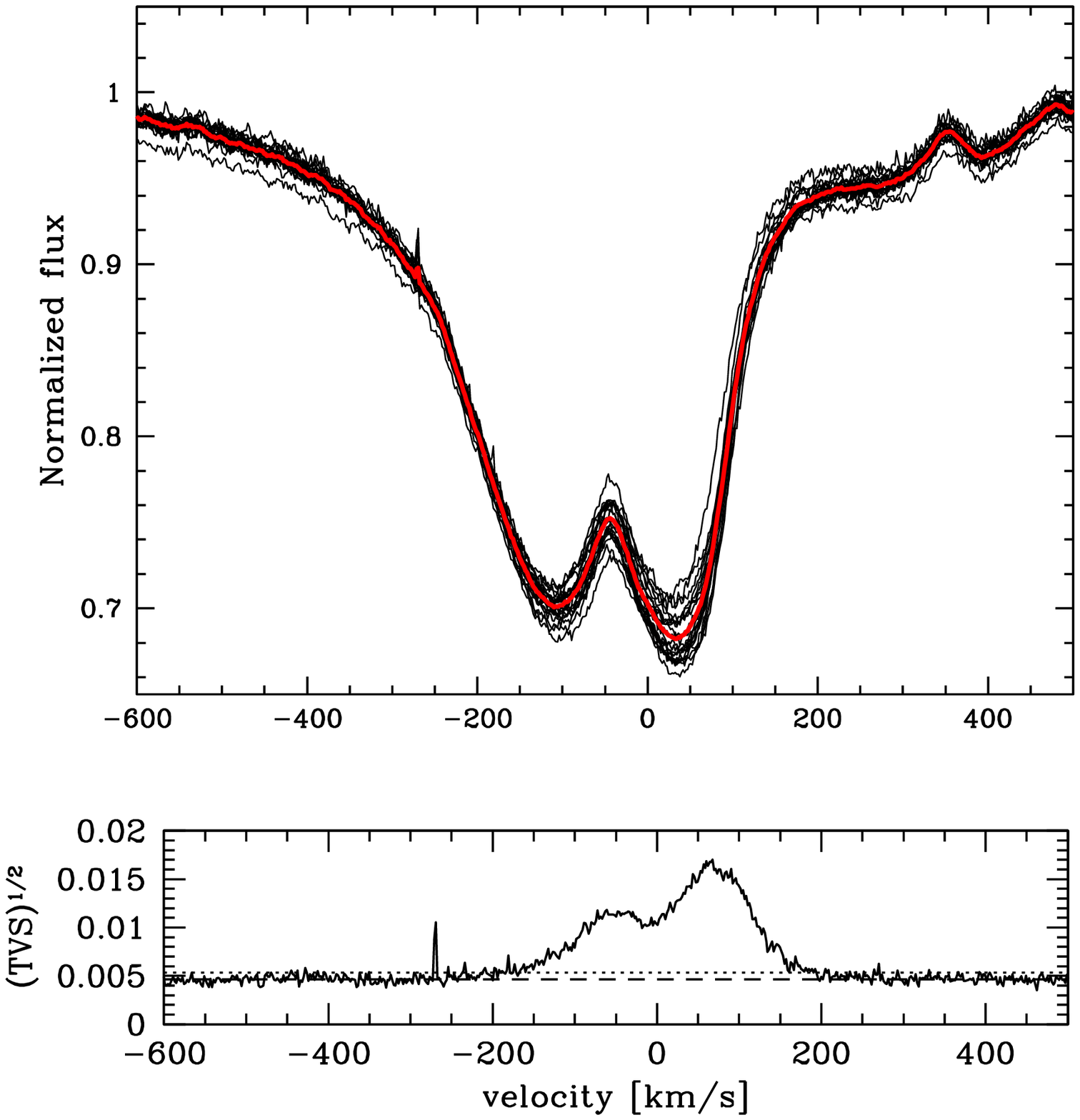}}\\
     \subfigure[He~{\sc i}~$\lambda$~4471]{
           \includegraphics[width=.31\textwidth]{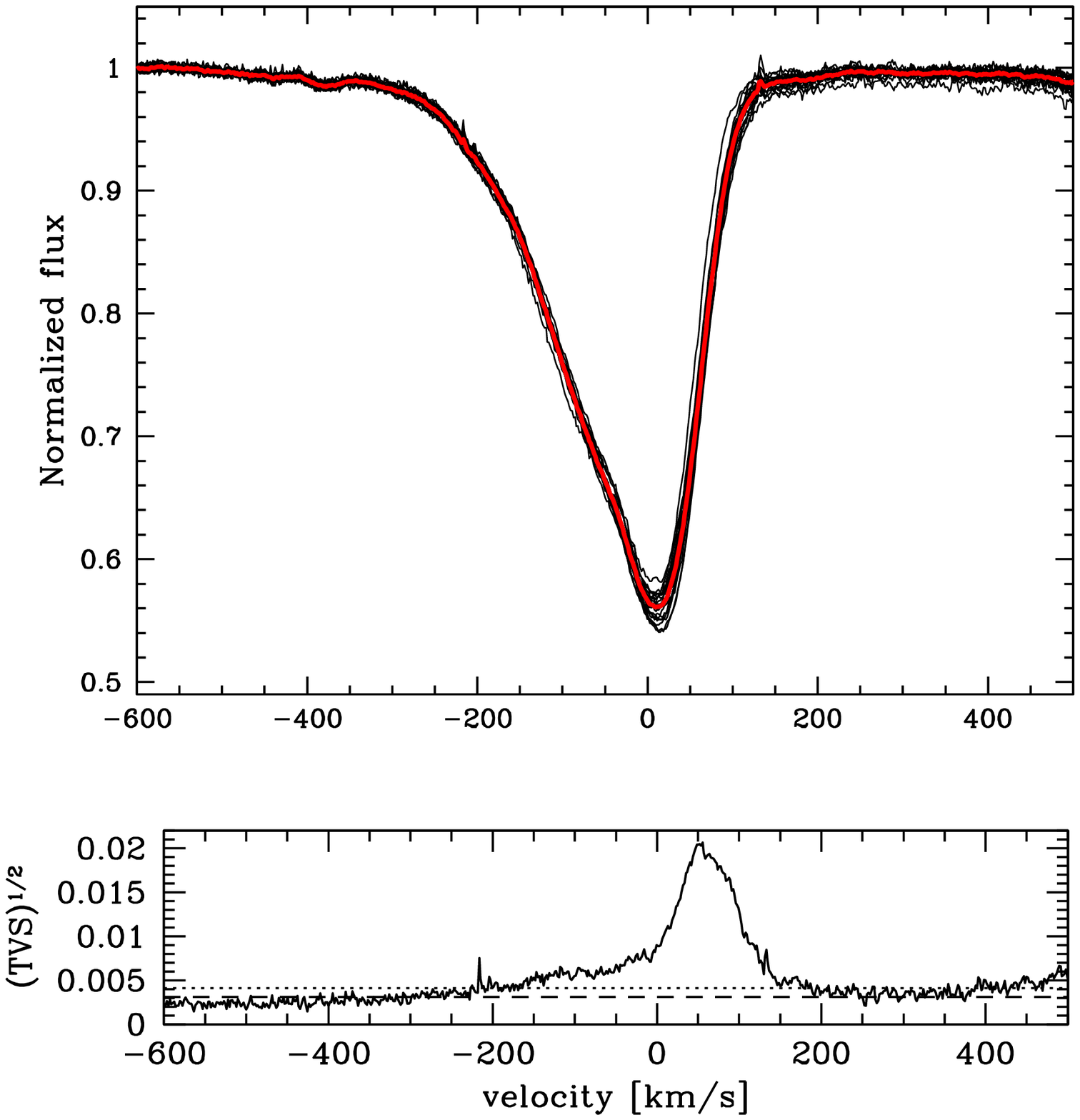}}
     \hspace{0.2cm}
     \subfigure[He~{\sc i}~$\lambda$~4713]{
          \includegraphics[width=.31\textwidth]{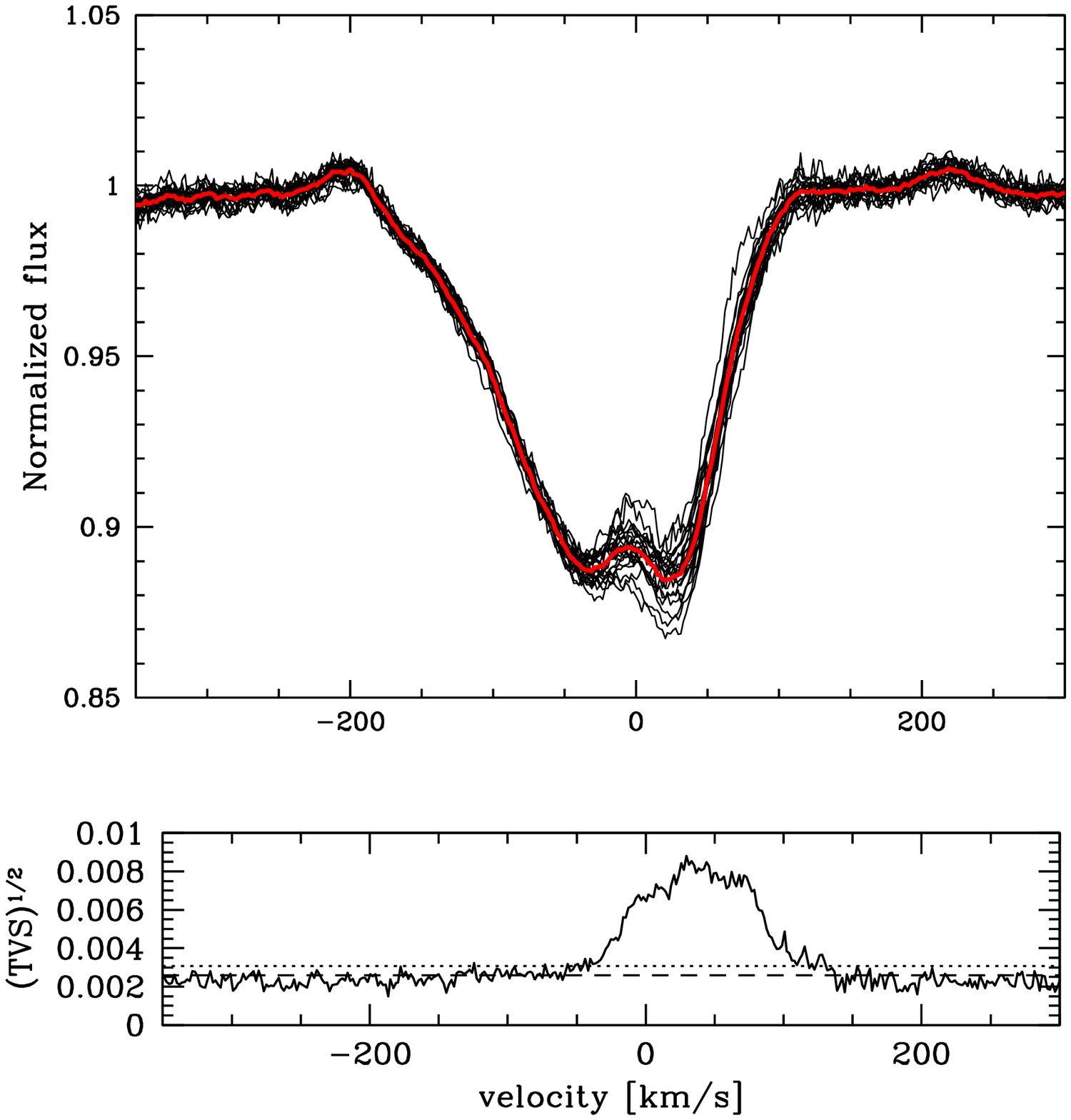}}
     \hspace{0.2cm}
     \subfigure[He~{\sc i}~$\lambda$~4920]{
          \includegraphics[width=.31\textwidth]{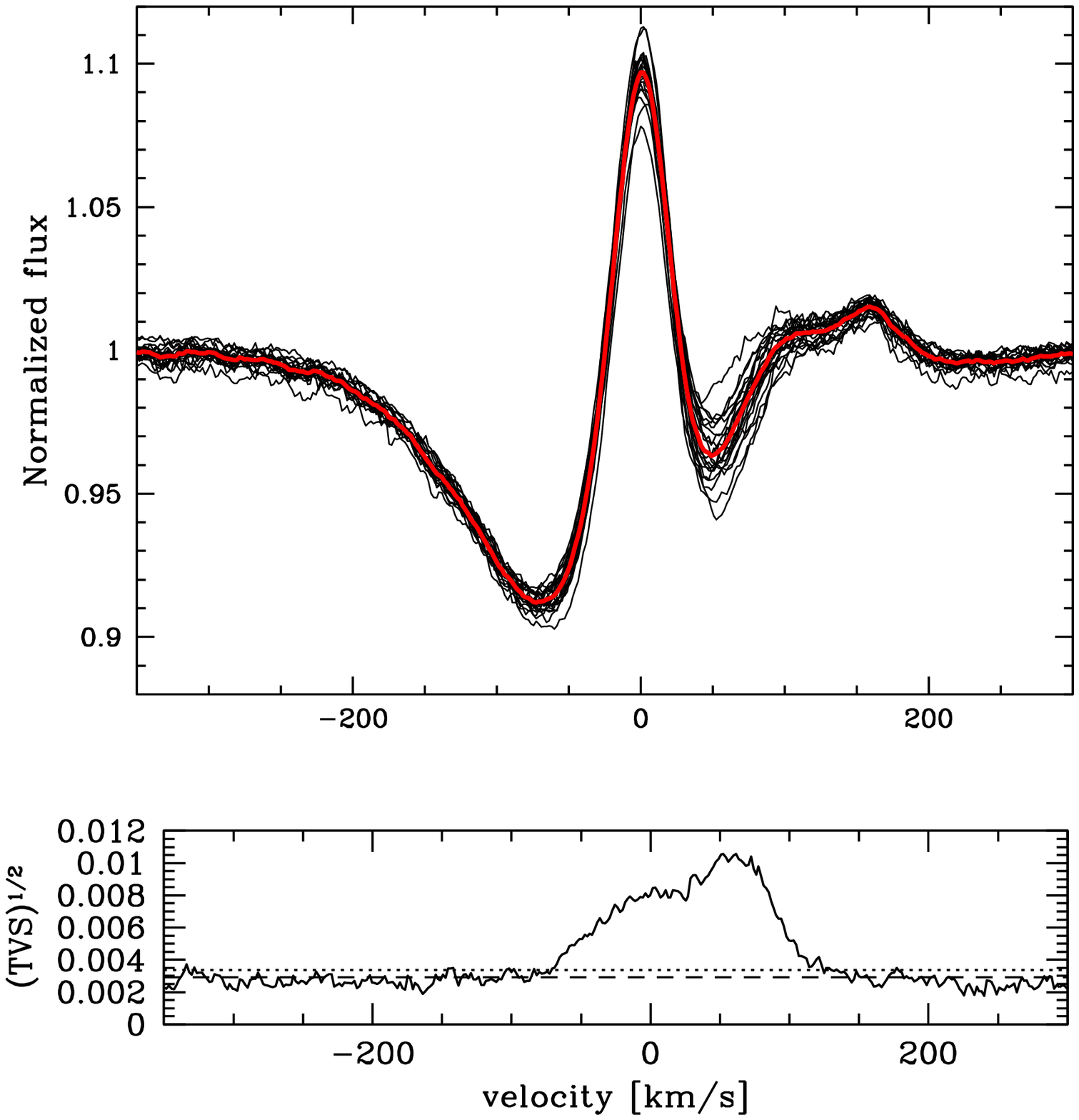}}\\
     \subfigure[He~{\sc ii}~$\lambda$~4200]{
           \includegraphics[width=.31\textwidth]{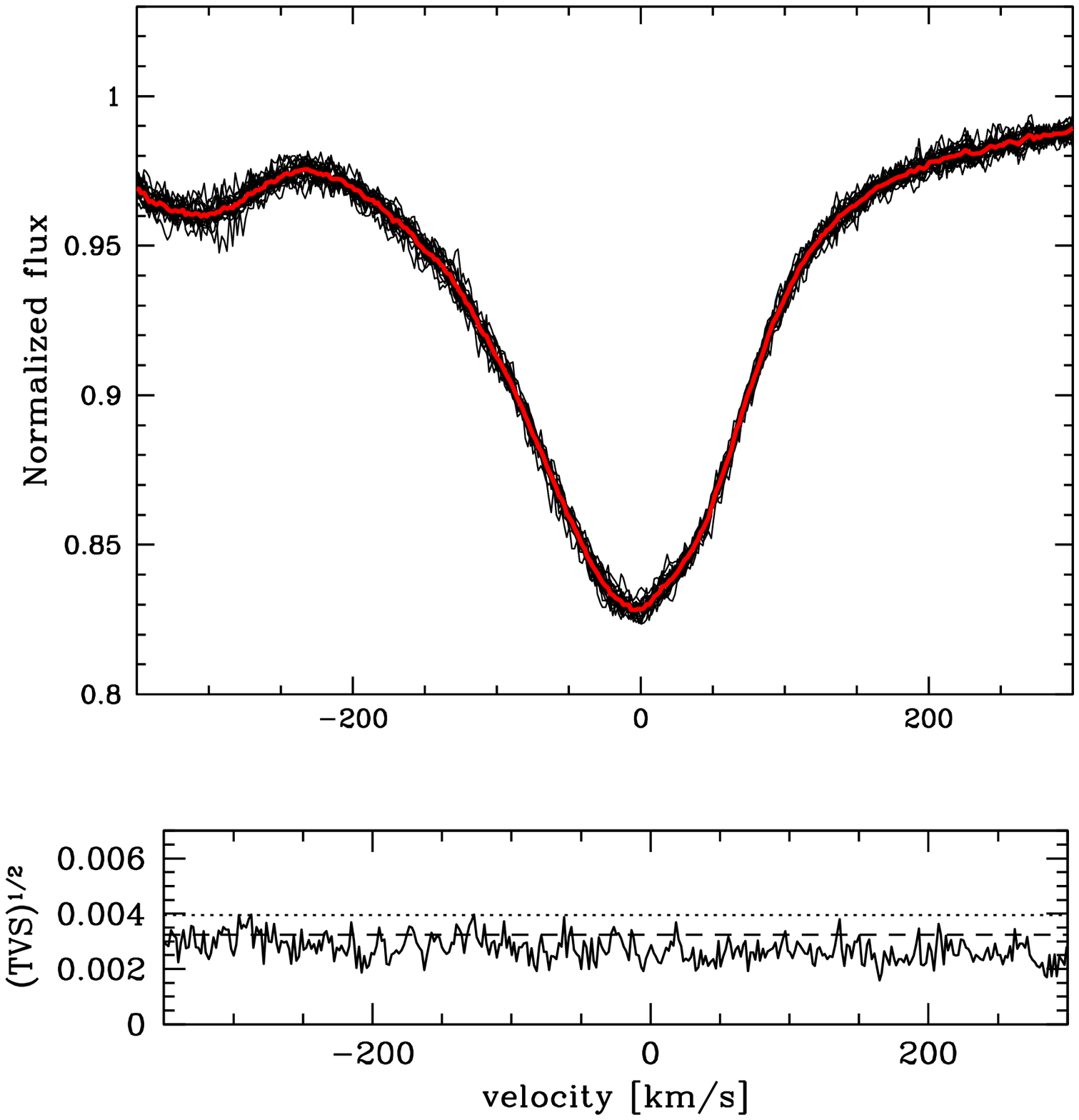}}
     \hspace{0.2cm}
     \subfigure[He~{\sc ii}~$\lambda$~4542]{
          \includegraphics[width=.31\textwidth]{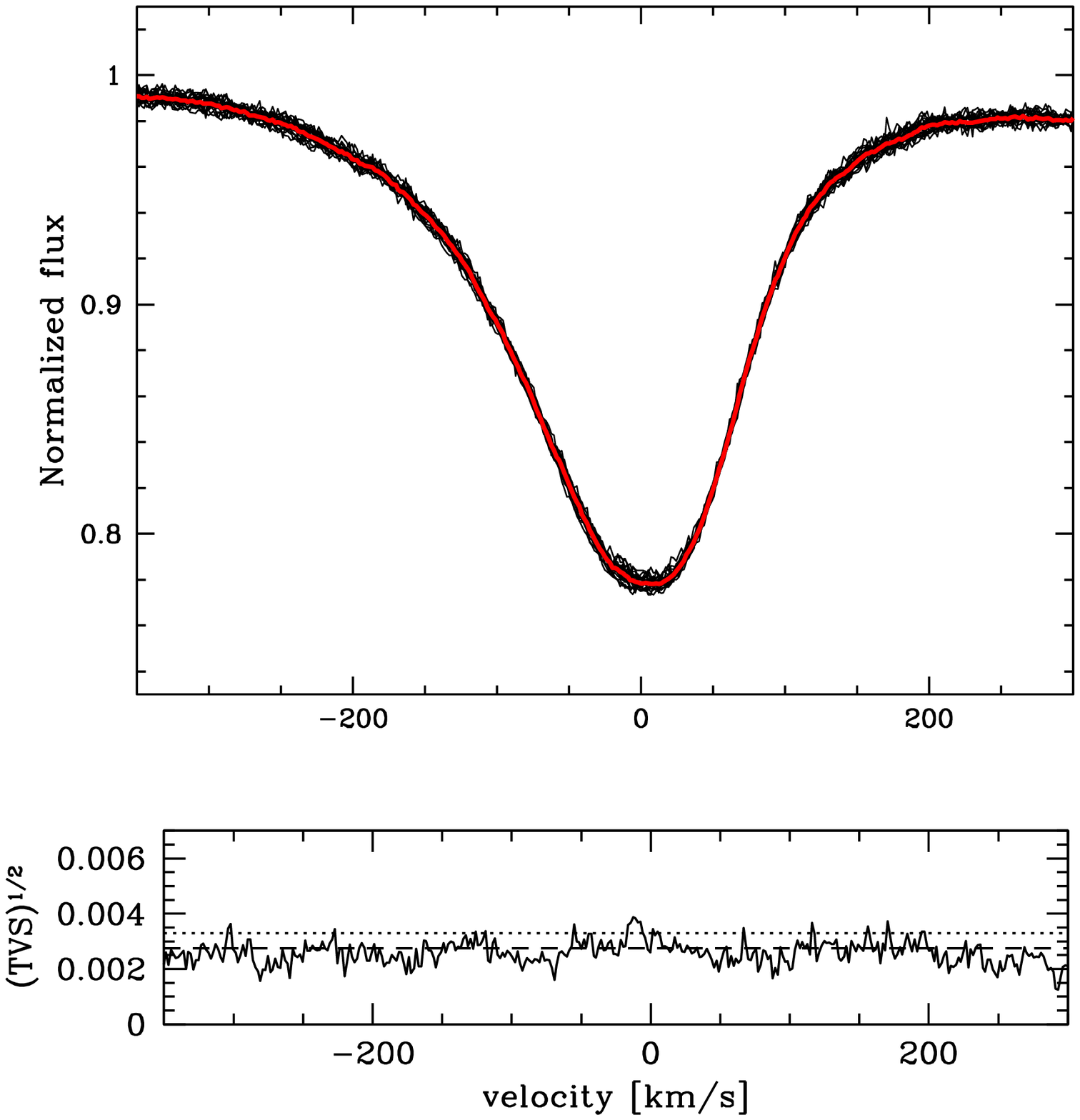}}
     \hspace{0.2cm}
     \subfigure[He~{\sc ii}~$\lambda$~4686]{
          \includegraphics[width=.31\textwidth]{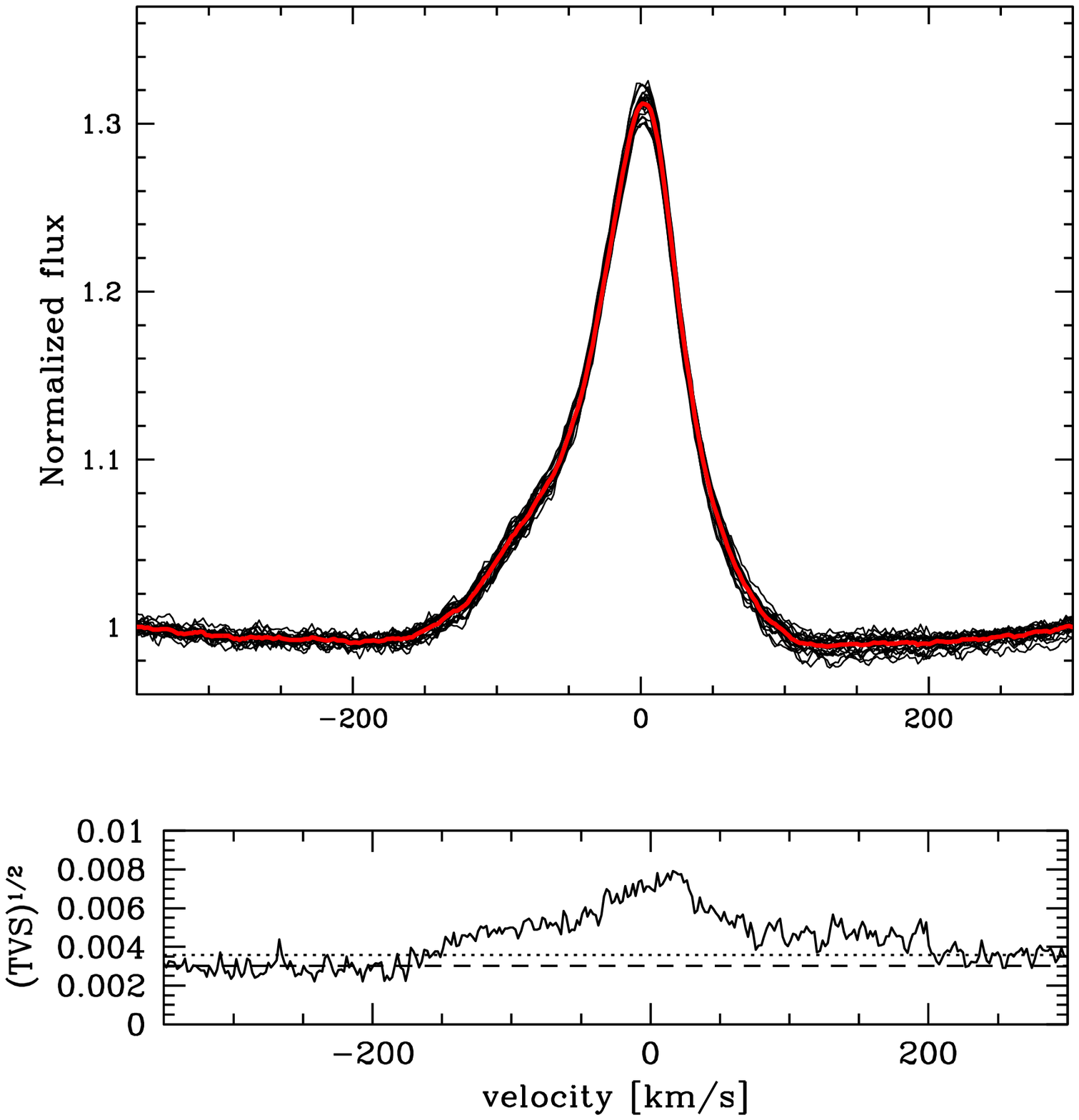}}\\
     \caption{Line profile variability of H$\alpha$, H$\beta$, H$\gamma$,  He{\sc i}~$\lambda$~4471,  He{\sc i}~$\lambda$~4713, He{\sc i}~$\lambda$~4920,  He{\sc ii}~$\lambda$~4200, He{\sc ii}~$\lambda$~4542 and He{\sc ii}~$\lambda$~4686 (panels a to i) between July 4$^{th}$ and October 9$^{th}$ 2009.  The red bold line is the 2009 average spectrum while the thin lines correspond to individual spectra. All spectra are shown in velocity space around the respective rest wavelength.  The temporal variance spectrum \citep[TVS, see][]{full96} is also given for each line (lower panels). The dot-dashed line shows the one $\sigma$ deviation of the continuum, while the dotted lines shows the 99$\%$ confidence limit. Variability is clearly detected in all Balmer and He{\sc i} lines. On the contrary, the He{\sc ii} lines are stable, with only a small variability in the wind line He{\sc ii}~$\lambda$~4686.}
     \label{tvs}
\end{figure*}

The spectrum of HD~108 is varying on timescales of a few decades. Our excellent time coverage in 2009 allows us to investigate the existence of faster variability. Fig.\ \ref{tvs} shows a selection of He~{\sc i}, He~{\sc ii} and Balmer lines together with the corresponding Time Variance Spectrum \citep[TVS, see][]{full96} which quantifies the degree of variability. All He~{\sc i} and Balmer lines are variable on timescales of days. He~{\sc ii}~$\lambda$~4686 also shows fluctuations. But the He~{\sc ii}~$\lambda$~4200 and He~{\sc ii}~$\lambda$~4542 lines (as well as C~{\sc iv}~$\lambda\lambda$~5801--5812 and O~{\sc iii}$\lambda$5592, not shown in Fig.\ \ref{tvs}) are stable over time. The He~{\sc i} and Balmer lines, as well as the wind line He~{\sc ii}~$\lambda$~4686, are formed over a large fraction of the atmosphere (i.e. photosphere + wind), especially if a disk/equatorial density enhancement exists. On the contrary, He~{\sc ii}~$\lambda$~4200 and He~{\sc ii}~$\lambda$~4542 are formed in the hottest part of the atmosphere, in or very close to the photosphere. The observed variability thus originates in the wind. Non-radial pulsations can thus be excluded since they would trigger variability in photospheric lines. A periodicity search performed on equivalent width measurements of several lines and using the Lomb-Scargle formalism did not reveal any pattern. We thus conclude that HD~108 presents day-to-day stochastic wind variability. 

In the standard magnetically confined wind scenario, and in the case of moderate to strong confinement, a disk is formed in the inner atmosphere \citep[e.g. Fig.\ 3 of][]{ud02}. This disk is denser in its inner parts. \citet{ud02} and \citet{ud06} have shown that in absence of sufficient centrifugal support, material accumulated in the disk and located below the co-rotation radius falls back onto the stellar surface \citep[see Fig.\ 4 of][]{ud02}. This process is stochastic, and the disc in maintained dynamically by new material injected via the channelled wind. In most lines of Fig.\ \ref{tvs} we see that variability is located preferentially on the red part of the profile. This can be interpreted as a direct evidence for the infall of ``blobs'' of disk material. Indeed, density fluctuation along the line of sight will cause variation of the line strength, and the receding velocity will shift these variations towards longer wavelengths. We think that this variability in the red wings of most line profiles is an indication of material infall. Similar conclusions were reached by \citet{wade06} for \tori. H$\alpha$ seems to deviate from this scenario: its variability is located over the entire profile. Contrary to most lines of Fig.~\ref{tvs}, H$\alpha$ is formed further away from the photosphere and is usually more affected by the wind. It is known to be variable in a number of non--magnetic O stars, mainly due to the presence of inhomogeneities (``wind--clumping'') most likely due to hydrodynamical instabilities \citep[e.g.][]{runacres02}. \citet{markova05} showed that such clumps could explain the H$\alpha$ variability observed over the entire line profile of O supergiants. We thus suggest that in HD~108, H$\alpha$ is formed in a zone where the putative disk of confined material is more tenuous than in the formation region of all other lines of Fig.\ \ref{tvs}. In that region, wind--clumping affects the line profile in addition to material infall. This explains that variability is not confined to the red wing but is located over the entire profile. The large clumping factor derived in Sect.\ \ref{s_wind} (f=0.01) is consistent with a very inhomogeneous wind. Obviously, more information on the field strength and geometry are needed to better constrain the wind confinement and to test our suggestion. In particular, a better knowledge of the position of the Alfv\'en radius relative to the H$\alpha$ formation region is important to test our hypothesis that H$\alpha$ variability is at least partly caused by wind--clumping. 

It is also worth noting that the models of \citet{ud02} and \citet{ud06} consider only the case of magnetic axis aligned with the rotation axis. In our case a tilt is likely present given the long--term line variability, so that the spin--down timescale and centrifugal support cannot be strictly compared to the theoretical predictions.

We finally stress that in the case of magnetically confined winds, one might question the use of 1D atmosphere models to constrain the wind properties. However, it is unlikely that our mass loss rate estimate is wrong by orders of magnitudes. Indeed, even if half of the material initially blown by radiative acceleration was to fall back onto the stellar surface, strong signatures of infall such as inverse P--Cygni profiles or redshifted absorption should be seen. Since the only evidence for infall is the line variability of absorption lines, we think our estimate of the mass loss rate is realistic.

\section{SUMMARY AND CONCLUSIONS}
\label{s_conc}

We have presented spectropolarimetric observations of the Of?p star
HD~108 conducted with the NARVAL and ESPaDOnS instruments (at TBL and
CFHT respectively). One hundred and ten circularly--polarized spectral sequences have been collected between 2007 and
2009. We report the clear detection of a Stokes V Zeeman signature
stable on timescales of days to months, but likely slowly increasing in
amplitude on timescales of years. We speculate that this timescale is
the same as that on which the H$\alpha$ emission is varying and is equal
to the rotation period of the star. The corresponding longitudinal
magnetic field is of order of 100--150~G, implying that the polar
strength of the putatively-dipolar large-scale magnetic field of
HD~108 is at least 0.5~kG and most likely of order of 1--2 kG.

The stellar and wind parameters of HD~108 have been derived through atmosphere modelling with the code CMFGEN. The effective temperature was poorly constrained due to the surprisingly strong He~{\sc i}~$\lambda$~4471 and He~{\sc i}~$\lambda$~5876 lines. Values of \teff\ in the range 33000--37000 K were conservatively derived from the analysis of the remaining He~{\sc i} and the He~{\sc ii} lines. A mass loss rate of the order $10^{-7}$ \myr\ was derived. The wind is found to be strongly clumped (f=0.01). It is also significantly confined ($\eta_{\star} \geq 100$, possibly up to 800) by the magnetic field. HD~108 shows long-term variability in most He~{\sc i} and Balmer lines, with profiles changing from pure absorption to P--Cygni on a few decades period. We suggest that this periodicity corresponds to rotational modulation. The (magnetic) equatorial density enhancement implied by the wind confinement would then be seen edge-on at minimum line emission, and pole-on at maximum emission. The timescale for magnetic braking computed with the derived stellar, wind and magnetic properties is consistent with the star's age. A slow rotation is also consistent with the low \vsini\ we derive. HD~108 might thus be an even more extreme case of slowly rotating magnetic O star than HD~191612.

Short-term line profile variability is also observed in He~{\sc i} and wind sensitive lines. Photospheric He~{\sc ii} lines are very stable. Lines formed just above the photosphere show evidence for infall while the variability of lines formed in the outer wind is more typical of normal hot stars' outflows. We suggest that in the inner wind, we are witnessing the infall of material channeled by the field lines to the magnetic equator and subsequently pulled back to the stellar surface by gravity.

Most of the suggestions discussed in Sect.\ \ref{s_disc} and summarized above are still speculative. Future monitoring of HD~108 is obviously needed to confirm the expected correlation between the variation of the longitudinal magnetic field and the long-term spectroscopic variability. If the few decades spectroscopic and photometric modulation corresponds to the rotational period, a complete mapping of the magnetic topology is not possible until a few decades. However, crucial information regarding the field strength and geometry can be gathered in the next years since we have just passed the phase of minimum emission. The field is thus expected to strengthen, making detection easier. Further constraints on the field morphology will help to see 1) if the slow rotation of HD~108 is really a due to magnetic braking and 2) if the short-term variability is caused by dynamical phenomena predicted by current simulations of magnetically--channelled winds.

\section*{Acknowledgments} 
We thank John Hillier for making his code CMFGEN available and for constant help with it.
We also thank the generous time allocation from the TBL TAC and
the MagIcS initiative under which MiMeS is carried out. We acknowledge the help of the
TBL and CFHT staff for service and QSO observing respectively. GAW acknowledges Discovery Grant support from the Natural Science and Engineering Research Council of Canada (NSERC). JCB and WM acknowledge financial support from the French National Research Agency (ANR) through program number ANR-06-BLAN-0105.

\bibliography{hd108.bib}

\bsp 
 
\label{lastpage} 
 
\end{document}